\documentclass[amsmath,amssymb,prl,hyperlink,twocolumn,showpacs]{revtex4}

\usepackage{graphicx}
\usepackage{soul}
\usepackage[colorlinks=true,citecolor=blue,linkcolor=magenta]{hyperref}
\usepackage[usenames]{color}
\usepackage{amsfonts}
\usepackage{color}
\usepackage{booktabs}
\usepackage{multirow}
\usepackage{float}
\usepackage[british]{babel}
\usepackage[style=american]{csquotes}
\usepackage{times}
\usepackage{epstopdf}

\MakeOuterQuote{"}

\begin{document}

\title{Genuine 12-qubit entanglement on a superconducting quantum processor}
\author{Ming Gong,$^{1,2}$ Ming-Cheng Chen,$^{1,2}$ Yarui Zheng,$^{1,2}$ Shiyu Wang,$^{1,2}$ Chen Zha,$^{1,2}$ Hui Deng,$^{1,2}$ Zhiguang Yan,$^{1,2}$ Hao Rong,$^{1,2}$ Yulin Wu,$^{1,2}$ Shaowei Li,$^{1,2}$ Fusheng Chen,$^{1,2}$ Youwei Zhao,$^{1,2}$ Futian Liang,$^{1,2}$ Jin Lin,$^{1,2}$ Yu Xu,$^{1,2}$ Cheng Guo,$^{1,2}$ Lihua Sun,$^{1,2}$ Anthony D. Castellano,$^{1,2}$ Haohua Wang,$^{3}$ Chengzhi Peng,$^{1,2}$ Chao-Yang Lu,$^{1,2}$ Xiaobo Zhu,$^{1,2}$ and Jian-Wei Pan$^{1,2}$\vspace{0.2cm}}

\affiliation{$^1$ Hefei National Laboratory for Physical Sciences at Microscale and Department of Modern Physics, University of Science and Technology of China, Hefei, Anhui 230026, China}
\affiliation{$^2$ CAS Centre for Excellence and Synergetic Innovation Centre in Quantum Information and Quantum Physics, University of Science and Technology of China, Hefei, Anhui 230026, China.}
\affiliation{$^3$ Department of Physics, Zhejiang University, Hangzhou, Zhejiang 310027, China}
\date{\today}

\begin{abstract}
	
	We report the preparation and verification of a genuine 12-qubit entanglement in a superconducting processor. The processor that we designed and fabricated has qubits lying on a 1D chain with relaxation times ranging from 29.6 to 54.6 $\mu$s. The fidelity of the 12-qubit entanglement was measured to be above $0.5544\pm0.0025$, exceeding the genuine multipartite entanglement threshold by 21 {statistical} standard deviations. Our entangling circuit to generate linear cluster states is depth-invariant in the number of qubits and uses single- and double-qubit gates instead of collective interactions. Our results are a substantial step towards large-scale random circuit sampling and scalable measurement-based quantum computing.
	
\end{abstract}

%\keywords{}

\maketitle

\begin{figure*}[t]
	\centering
	\includegraphics[width=0.7\textwidth]{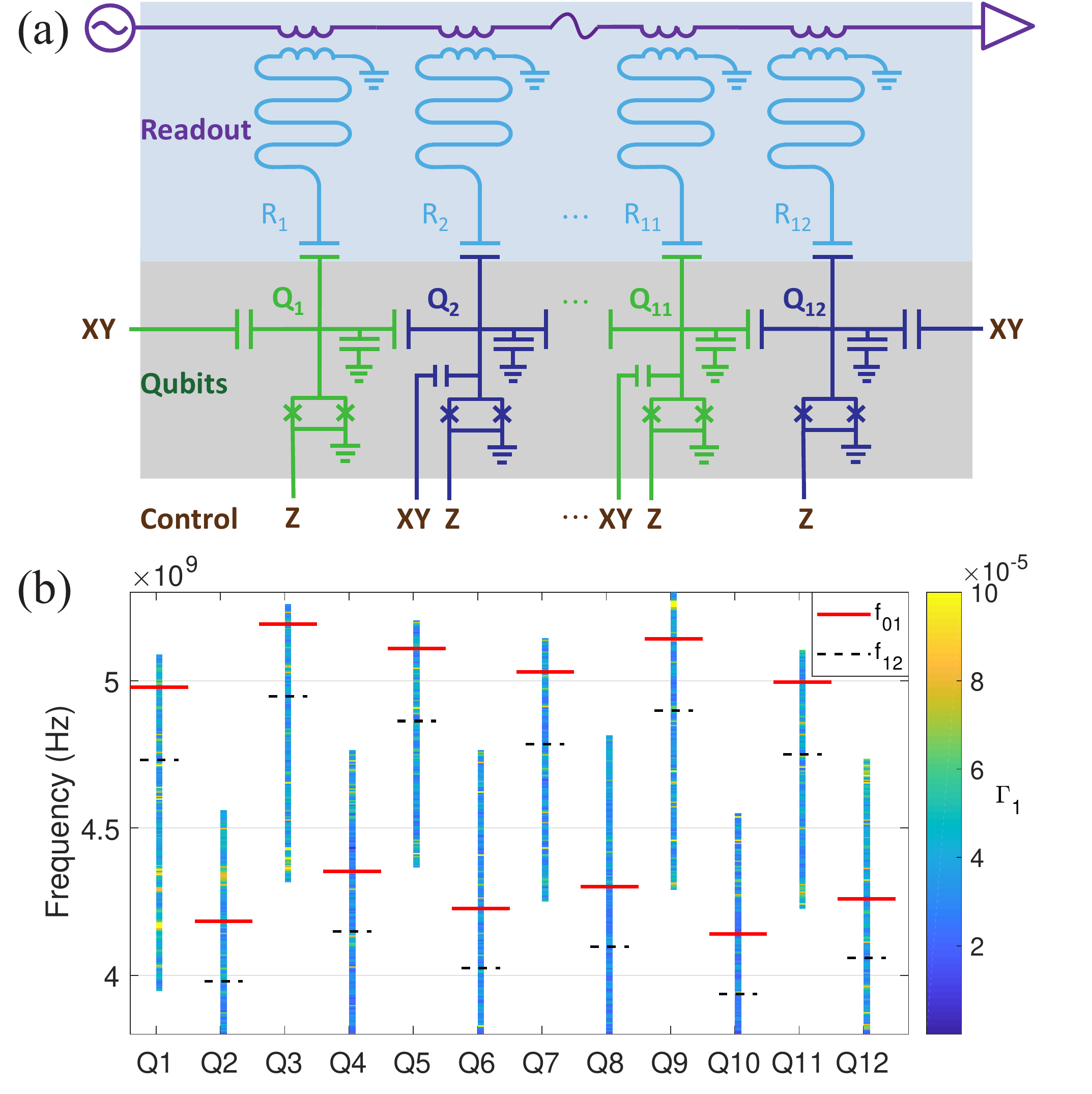}
	\caption{(a) Circuit diagram. There are 12 neighbouring qubits illustrated in two colours (green/dark-blue) which correspond to two groups of working frequencies. The green ones are around $5$ GHz and the dark-blue ones are around $4.2$ GHz. All readout resonators (light-blue) are coupled to a common transmission line (purple). By using frequency-domain multiplexing, joint readout for all qubits can be performed. For each qubit, individual capacitively-coupled microwave control lines (XY) and inductively-coupled bias lines (Z) enable full control of qubit operations. (b) The idling frequencies of both $f_{01}$ (solid red line) and $f_{12}$ (dotted black line) for all qubits. The colour of the vertical bars on qubit levels indicate the energy relaxation rate $\Gamma_1$. All qubit operations are performed within this frequency range. }
	\label{fig1}
\end{figure*}

\begin{figure*}[t]
	\centering
	\includegraphics[width=\textwidth]{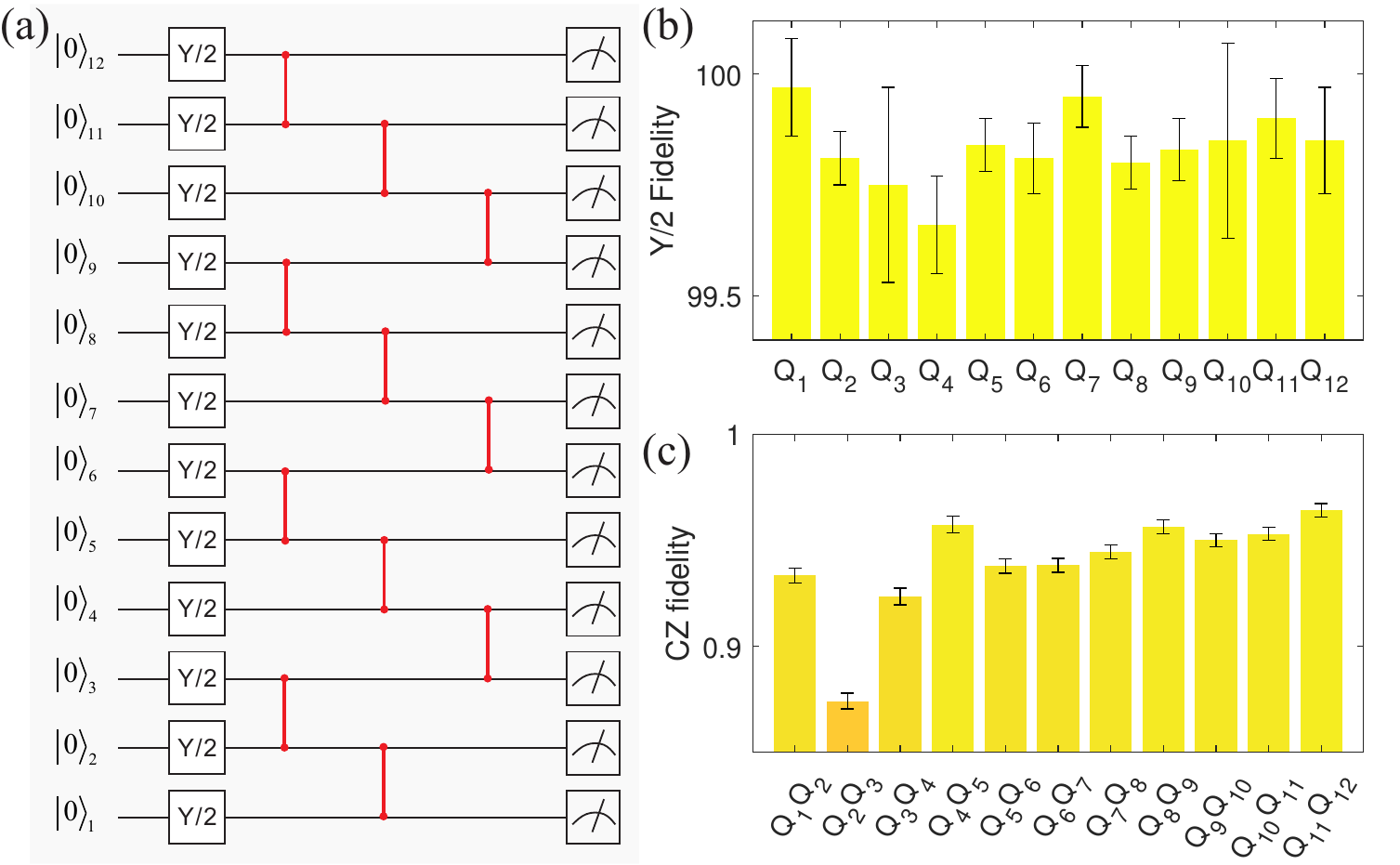}
	\caption{(a) Algorithm to generate linear cluster state. The initial state for each qubit is $|0\rangle$. $Y/2$ gates are applied to bring each qubit to $|+\rangle$, then the CZ gates are applied to generate the GME state. Finally, joint measurements are performed to obtain the state fidelity. (b) The fidelities of $Y/2$ single qubit gates obtained by randomized benchmarking. {The error bars represent a 95\% confidence interval, determined from nonlinear least-squares fits.} (c) {CZ gate fidelities obtained by performing quantum process tomography. Error bars on the data are calculated via bootstrapping method, with a 95\% confidence interval.}}
	\label{fig2}
\end{figure*}

\begin{figure*}[t]
	\centering
	\includegraphics[width=\textwidth]{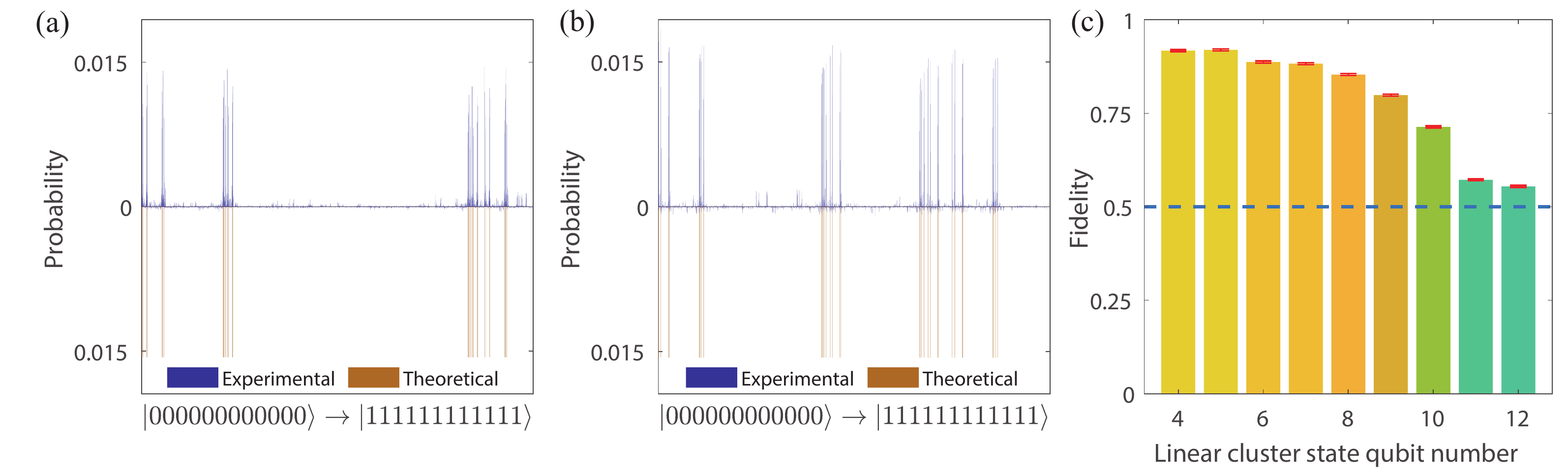}
	\caption{(a) Experimental and theoretical distribution of $XZ...XZ$ component in 12 qubit linear cluster state. (b) Same as (a), but use ZX...ZX instead. In both (a) and (b), the states are in the form of $|Q_{12}...Q_{1}\rangle$. Experimental and theoretical results are presented in dark-blue and brown, respectively.  (c) Linear cluster state fidelities from 4 to 12 qubits, which are 0.9176(28), 0.9196(28), 0.8870(27), 0.8827(27),
		0.8536(27), 0.7988(27), 0.7136(26), 0.5720(25) and 0.5544(25), exceed GME threshold by 149, 149, 143, 141, 130, 110, 82, 28 and 21 {statistical} standard deviations, respectively. State fidelities are calculated from two components, $XZ...XZ$ and $ZX...ZX$, of the linear cluster state. Error bars have a confidence interval of 95\%, obtained from statistical calculation. A threshold of 50\% for genuine entanglement is marked with a blue dashed line. }
	\label{fig3}
\end{figure*}

Quantum entanglement is a highly nonclassical aspect of quantum mechanics \cite{entanglement09, testingfundations14}, and a central resource to quantum information sciences \cite{computing10, simulating09, internet08, metrology11}. A stringent benchmark for high-precision control of multiple quantum systems is the ability to create genuine multipartite entangled (GME) state that cannot be expressed as a biseparable state or mixture of biseparable states with respect to variable partitions \cite{witness09}. So far, GME states in the form of Greenberger-Horne-Zeilinger (GHZ) states have been reported with 10 superconducting qubits \cite{10xmons}, 14 trapped ions \cite{14ions}, and 18 photonic qubits \cite{18GHZ}. We note that in several other experiments involving large numbers of qubits \cite{20ions,53ions,51Atoms,512spins,16ibm}, the presence of genuine entanglement for more than 5 qubits has not been verified. Here, we report the creation and verification of a 12-qubit linear cluster (LC) state, the largest GME state reported in solid-state quantum systems. LC states are robust against noise, and serve as a universal resource for one-way quantum computing \cite{ oneway01, oneway09}. Our approach does not rely on collective interactions to create GME as in the previous work \cite{10xmons,14ions}, but is based on individual single-qubit gates and controlled-phase (CZ) entangling gates, which makes our approach scalable to larger numbers of qubits and applicable to random quantum circuit sampling demonstrations of quantum supremacy \cite{sampling}.

 An $N$-qubit cluster state is a simultaneous eigenstate of $N$ commuting Pauli stabilizer operators with eigenvalues all equal to $+1$ \cite{oneway01}. Stabilizer operators consist of nearest-neighbour interactions of qubits arranged in lattices. The simplest example is a linear cluster (LC) state, where stabilizer operators ${s_i}$  are defined on a qubit chain as
\begin{equation}
{s_i} = \sigma _Z^{(i - 1)}\sigma _X^{(i)}\sigma _Z^{(i + 1)}
\end{equation}
$\sigma _X^{(i)}$ and $\sigma _Z^{(i)}$ are Pauli X and Z operators on $i$-th qubit, respectively (and at the boundary $\sigma _Z^{(0)}$ and $\sigma _Z^{(N+1)}$ are idle). Cluster states can be prepared either by cooling a nearest-neighbour Ising-type Hamiltonian   $H = \sum\nolimits_{i = 1}^N {\frac{{1 - {s_i}}}{2}} $ system to its ground state or by dynamically implementing a set of CZ gates
\begin{equation}
	| {L{C_N}} \rangle  = \bigg[ \prod _{i = 1}^{N - 1}C{Z^{(i,i + 1)}} \bigg] {| + \rangle ^{ \otimes N}} \label{2}
\end{equation}
on a qubit lattice initialized in the $\left|  +  \right\rangle  = (\left| 0 \right\rangle  + \left| 1 \right\rangle )/\sqrt 2 $ state. In this work, we use the latter method on a superconducting quantum processor by implementing the gate sequence shown diagrammatically in Fig. \ref{fig2}(a).

As can also be seen in Fig. \ref{fig1}(a), the processor has 12 transmon qubits \cite{tranmon07} of the Xmon variety \cite{RB14}. Each qubit has a microwave drive line (XY), a fast flux-bias line (Z) and a readout resonator. The qubits are arranged in a line with neighbouring qubits  coupled capacitively. All the readout resonators are coupled to a common transmission line for joint readout of the qubit states. The Hamiltonian of the 12-qubit system is given by
\begin{equation}
H/\hbar=\sum_{i=1}^{12}\omega_i\hat{n}_i+\frac{\eta_i}{2}\hat{n}_i(\hat{n}_i-1)+\sum_{i=1}^{11}g_i(\hat{a}_i^\dagger\hat{a}_{i+1}+\hat{a}_i\hat{a}_{i+1}^\dagger)
\end{equation}
where $\hat{n}$ is the number operator, $\hat{a^\dagger}$ ($\hat{a}$) is the creation (annihilation) operator, $\omega_i$ and $\eta_i$ are the transition frequency and the anharmonicity of the $i$-th qubit, respectively, and $g_i$ is the interaction strength between $i$-th and $(i+1)$-th qubits. Each qubit transition frequency can be tuned by Z lines and single-qubit quantum gates can be implemented by driving the XY lines. For specific qubit properties, refer to the supplemental information \cite{seesm}.

The specific quantum circuit used to produce the LC state is illustrated in Fig. \ref{fig2}(a). To perform the entire operation, first we wait 300 $\mu s$ to relax the qubits into the $|0\rangle$ state. Then, we apply $Y/2$ gates to rotate all the qubits into the $|+\rangle$ state. After that, 11 CZ gates are performed to entangle all 12 qubits. Finally, we measure all qubit states with a joint readout operation.

The nearest-neighbour coupling enables the application of "fast adiabatic" CZ gates \cite{RB14}. To minimize the effects of decoherence and $ZZ$ coupling between neighbouring qubits, we shorten the depth of the circuit by applying the CZ gates in parallel. {The minimization of $ZZ$ coupling also requires a large detuning between adjacent qubits. We carefully arranged the idle frequencies to avoid TLSs and adjust the frequency differences between adjacent qubits larger than $700$ MHz. The idle frequencies for all relevant qubits are shown in Fig. \ref{fig1}(b).} Choosing a gate sequence like this, along with carefully optimizing and calibrating the control pulses, was crucial to achieve this high fidelity entanglement. We have put the relevant technical details into the supplemental information \cite{seesm}.

  %Note that when we characterize a single entangling gate, we run the entire three-layer sequence, so that the final fidelity can reach its maximum (see supplemental information for details).

{The fidelities of the $Y/2$ gates and CZ gates are reported in Fig. \ref{fig2}(b) and (c), respectively. The fidelities of CZ gates} are calculated using quantum process tomography (QPT), where maximum-likelihood estimation is used to construct physical density matrices resulting from an arbitrary input. The average CZ gate fidelity is 0.939. But it is also possible to characterize our gates for states initialized in $|{+}{+}\rangle$, in which case the average fidelity increases to 0.956. This is more relevant to our experiment because CZ gates are only ever applied to the $|{+}{+}\rangle$ state. The $Q_2$-$Q_3$ gate is the worst of all the CZ gates. This is caused by defects in the physical system located on $Q_3$ around 4.43 GHz and on $Q_2$ around 4.34 GHz, which appear in Fig. \ref{fig1}(b) as a dramatic increase of the relaxation rate in a narrow range of frequencies. These so-called two-level systems (TLS) cause a qubit state to leak out of the computational state space, limiting the gate fidelity. Ignoring $Q_2$ and $Q_3$, the rest of the qubits have an average gate and state fidelity which increase to 0.946 and 0.962, respectively.

The fidelities of the CZ gates characterized here are lower than the actual gate fidelities. This is partly because unlike randomized benchmarking (RB), our characterization process includes errors from state preparation and readout. Also, when we characterize a single entangling gate, we run the entire three-layer sequence, which makes the effects of decoherence and $ZZ$ coupling larger due to the tripled length of the operation ($192$ ns). Fidelities of a single CZ gate for this processor, characterized by RB, typically exceed $0.99$. However, optimizing the CZ gates by embedding them into the whole circuit is essential, otherwise a high-fidelity GME state is unobtainable.

Fidelity measurements of states produced in quantum information experiments are traditionally calculated from the state's density matrix, which is obtained from quantum state tomography (QST). This full characterization of a state requires measurements and computational resources that grow exponentially in the number of qubits. In this work, full characterization proves impractical, so we find a lower bound of the state fidelity using
\begin{equation}
\mathcal{F}\ge  \vec{\alpha}_{XZ} \cdot {P}_{XZ}  + \vec{\alpha}_{ZX} \cdot {P}_{ZX} - 1
\end{equation}
where $P_{XZ}$ and $P_{ZX}$ are probability distributions measured with $\sigma_{XZ...XZ}$ and $\sigma_{ZX...ZX}$ bases, and $\vec{\alpha}_{XZ}$ and $\vec{\alpha}_{ZX}$ are two sets of coefficients equal to the theoretical distribution times $2^6$ ($N=12$) \cite{witness05,witness09,witness11prl}. See the supplemental information \cite{seesm} for the justification of this bound.

The measured probability distributions, $P_{XZ}$ and $P_{ZX}$, of the 12-qubit cluster state are shown in Fig. \ref{fig3}(a)(b), along with the theoretical distributions of the ideal state. The infidelity can be calculated from the sum the small components in the measured $P_{XZ}$ and $P_{ZX}$ distributions. These values, which mainly come from the population imbalance and phase errors in the prepared states, don't interfere destructively in the measurement process.

In our experiments, the readout is a positive-operator valued measurement, and we use calibrated transition matrices to obtain the original distributions. Due to statistical fluctuations, small probability values may become negative. We note that the fidelity bound in this process is reliable (see the supplementary information \cite{seesm}). We perform 250,000 projective measurements to construct the probability distributions $P_{XZ}$ and $P_{ZX}$. The lower bound of the 12-qubit linear cluster state fidelity is calculated to be $0.5544\pm0.0025$. We also prepared other linear cluster states from 4 qubits to 11 qubits by initializing $N$ neighbouring  qubits in $|+\rangle$ states and leaving the other qubits in $|0\rangle$ states. The fidelities are summarized in Fig. \ref{fig3}(c).

Once we have a lower bound of the state fidelity, we use entanglement witness to prove that the prepared states are genuinely entangled \cite{witness09}. An arbitrary quantum state $\rho$ that is bisperarable will always have a fidelity
$ F = \rm Tr(\rho  \left| {L{C_n}} \right\rangle \left\langle {L{C_n}} \right|)$
less than 0.5, hence states with fidelity above 0.50 are genuinely entangled. Fig. \ref{fig3}(c) shows that all states produced meet this criterion for entanglement. For the case of 12 qubits, the fidelity is $0.5544\pm0.0025$ and exceeds the threshold for entanglement by 21 {statistical} standard deviations. We note that a reported 16-qubit "full entanglement" \cite{16ibm} is not necessarily a \textit{genuine multipartite entanglement} because it is possible to generate fully entangled states with classical mixtures of separable states. An example is given in the supplementary information \cite{seesm}.

Scalability is one of the key advantages of our system: any two linear cluster states can be combined to form a larger cluster state  by applying one additional CZ gate. Additionally, a chain of N qubits (for $N>4$) will always take three layers of CZ gates to create an LC state, so negative effects from decoherence and $ZZ$ crosstalk won't be exacerbated by an increased circuit depth. We judge that using identical technology, a 20-qubit LC state could be created, if not for the presence of TLSs in the physical qubits. {On our system, after the TLSs coupled to $Q_2$ and $Q_3$ successfully removed by thermal cycling, the 12-bit LC state fidelity is improved to higher than 0.7. For more discussion of TLSs, see the supplemental information \cite{seesm}.} TLSs are the most immediate obstacle towards scaling to larger systems, and more work needs to be done investigating their physical origins and devising methods to mitigate their effects on superconducting quantum processors. 

The LC states produced in this work have immediate applications to near-term quantum supremacy experiments. Random quantum circuit sampling experiments typically use gate sequences that alternate between randomly chosen single-qubit gates and entangling gates on qubits arranged in 1D or 2D lattices \cite{sampling}. These gates are optimized using individually using RB \cite{RB08,RB14,RB17}, but simultaneous implementation of the gates causes them to interfere with each other. Instead of standard 1- and 2-qubit gates, the cluster state production sequences in this work can also be used as building blocks for random quantum circuit sampling. The techniques demonstrated in this work, those of optimization of simultaneous gates, are well suited to address similar challenges posed by the random circuit sampling experiments.

In general, cluster states have notable applications and advantages. The most interesting application is probably one-way quantum computing, where the most common starting state is the cluster state. The complex structure of the cluster state entanglement makes it possible to generate every quantum state \cite{oneway09}, which allows for further research in feed-forward operations \cite{feedforward07,feedforward08} and subsequent computations to be performed in a fault-tolerant way \cite{onewayTopological07}. Cluster states have the property that as the number of qubits increases, violation of the Bell inequalities increases exponentially \cite{nonlocal05}. Also, in noisy environments, the lifetime of entanglement is independent of the number of qubits, while for GHZ states, the lifetime approaches zero with increasing qubit number \cite{robust01,robust04}. This makes the cluster state worthy of more theoretical and experimental investigation.

We thank the Laboratory of Microfabrication, University of Science and Technology of China, Institute of Physics CAS and National  Center for Nanoscience and Technology for the support of the sample fabrication. This research was supported by the National Basic Research Program (973) of China under Grant No.~2017YFA0304300, the Chinese Academy of Science, Alibaba Cloud and Science and Technology Committee of Shanghai Municipality. X.-B. Zhu is supported by NSFC under Grants No. 11574380. H.-H Wang is supported by NSFC under Grants No.11434008.

%\bibliographystyle{apsrev4-1}
%\bibliography{12cluster}


%{\bf Author Contributions} X.-B. Zhu, C.-Y. Lu and J.-W. Pan conceived the research. M.-C. Chen, M. Gong, and X.-B. Zhu designed the experiment. Y.-R. Zheng, S.-Y. Wang and C. Zha designed the sample, under the guidance of X.-B. Zhu and H.-H. Wang. H. Deng, Z.-G. Yan and H. Rong prepared the sample, under the guidance of X.-B. Zhu. M. Gong, Y.-L. Wu, S.-W. Li and F.-S. Chen carried out the measurements, under the guidance of X.-B. Zhu. Y.-L. Wu developed the programming platform QOS for measurements. M.-C. Chen and M. Gong performed theoretical predictions and numerical simulations, advised by C.-Y. Lu. M. Gong, M.-C. Chen and Y.-W. Zhao analyzed the results, under the guidance of X.-B. Zhu and C.-Y. Lu. F.-T. Liang, J. Lin, Y. Xu, C. Guo, and L.-H. Sun developed Room temperature electronics equipment, under the guidance of C.-Z. Peng. All authors contributed to discussions of the results. M. Gong, M.-C. Chen, A. D. Castellano, C.-Y. Lu and X.-B. Zhu led the effort in the development of manuscript.

\end{document}

\beginsupplement
	
	\title{Supplementary Information for "Genuine 12-qubit entanglement on a superconducting quantum processor"}
	\author{Ming Gong,$^{1,2}$ Ming-Cheng Chen,$^{1,2}$ Yarui Zheng,$^{1,2}$ Shiyu Wang,$^{1,2}$ Chen Zha,$^{1,2}$ Hui Deng,$^{1,2}$ Zhiguang Yan,$^{1,2}$ Hao Rong,$^{1,2}$ Yulin Wu,$^{1,2}$ Shaowei Li,$^{1,2}$ Fusheng Chen,$^{1,2}$ Youwei Zhao,$^{1,2}$ Futian Liang,$^{1,2}$ Jin Lin,$^{1,2}$ Yu Xu,$^{1,2}$ Cheng Guo,$^{1,2}$ Lihua Sun,$^{1,2}$ Anthony D. Castellano,$^{1,2}$ Haohua Wang,$^{3}$ Chengzhi Peng,$^{1,2}$ Chao-Yang Lu,$^{1,2}$ Xiaobo Zhu,$^{1,2}$ and Jian-Wei Pan$^{1,2}$\vspace{0.2cm}}

\affiliation{$^1$ Hefei National Laboratory for Physical Sciences at Microscale and Department of Modern Physics, University of Science and Technology of China, Hefei, Anhui 230026, China}
\affiliation{$^2$ CAS Centre for Excellence and Synergetic Innovation Centre in Quantum Information and Quantum Physics, University of Science and Technology of China, Hefei, Anhui 230026, China.}
\affiliation{$^3$ Department of Physics, Zhejiang University, Hangzhou, Zhejiang 310027, China}

	\date{\today}
	
	\maketitle
	
	%\tableofcontents
		
	\section{Gate calibration and optimization}
\begin{figure}
	\centering
	\includegraphics[width=0.5\textwidth]{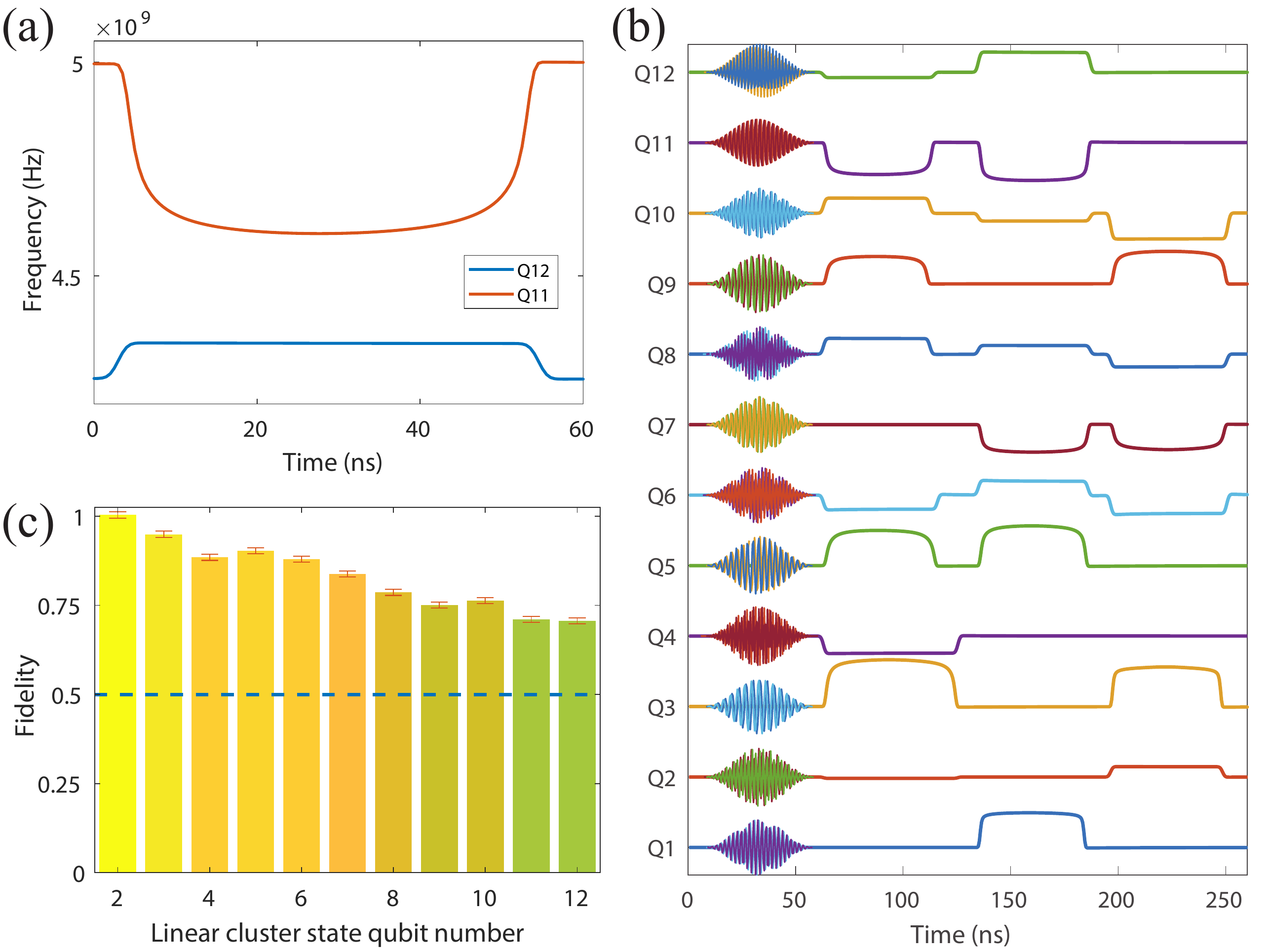}
	\caption{(a) Tuning trajectories in frequency of $Q_{12}$ and $Q_{11}$ in realizing a CZ gate. $Q_{12}$ is detuned from $4.258$ GHz to $4.343$ GHz, and $Q_{11}$ follows the "fast adiabatic" trajectory from $4.996$ GHz to $4.599$ GHz. (b) Waveform sequence in generating a 12-qubit linear cluster state. Parallel CZ gates are used to reduce total operation time. (c) {Linear cluster state fidelities from 2 to 12 qubits in the most recent experiment. After warming up the system to room temperature and cooling down again, the major TLSs coupled to $Q_2$ and $Q_3$ moved, which ensures the possibility in obtaining higher LC state fidelity. The state fidelities are 1.004(9), 0.949(9), 0.885(9), 0.903(9), 0.880(9), 0.838(9), 0.786(8), 0.750(8), 0.764(8), 0.711(8), and 0.707(8), exceed GME threshold by 56.3, 50.9, 44.3, 46.2, 43.8, 39.4, 33.9, 29.9, 31.4, 25.5, and 25.0 {statistical} standard deviations, respectively. For each state probability, we perform 25,000 projective measurement.}}
	\label{fig:s1}
\end{figure}

	The two-qubit CZ gate is implemented by tuning the $|11\rangle$ state close to the avoided crossing generated by the states $|11\rangle$ and $|02\rangle$ in an adiabatic trajectory, producing a state-dependent $\pi$ phase shift \cite{Strauch2003,Dicarlo2009,DiCarlo2010,Yamamoto2010,Kelly2014}. By controlling the adiabatic trajectory, one can suppress the leakage to the non-computational $|02\rangle$ state, while keeping the gate operation time relatively short. This is called a "fast adiabatic" trajectory.  For example, the CZ gate between $Q_{11}$ and $Q_{12}$ (see \ref{fig:s1}(a)) is implemented by tuning $Q_{11}$ from $4.996$ GHz down to $4.599$ GHz, and $Q_{12}$ from $4.258$ GHz up to $4.343$ GHz. The rising edge of $Q_{12}$ occurs just before $Q_{11}$ is modulated; symmetrically, the falling edge of $Q_{12}$ is placed slightly after that of $Q_{11}$. During this process, $Q_{10}$ is detuned down to avoid unwanted $ZZ$ coupling with $Q_{11}$. Since the anharmonicity of $Q_{11}$ was measured to be about -246 MHz, the $|11\rangle$ state is indeed brought close to the $|02\rangle$ state.

It takes a minimum of $N-1$ gates to create an N-qubit LC state. To reduce the total operation time, we applied the CZ gates in parallel. In our experiment, 11 CZ gates are sorted into 3 layers, $\{Q_{12}-Q_{11}$, $Q_9-Q_8$, $Q_6-Q_5$, $Q_3-Q_2\}$, $\{Q_{11}-Q_{10}$, $Q_8-Q_7$, $Q_5-Q_4$, $Q_2-Q_1\}$ and $\{Q_{10}-Q_9$, $Q_7-Q_6$, $Q_4-Q_3\}$. All CZ gates in one layer are applied in parallel. In this way, we reduce the total CZ operation time by a factor of $(N-1)/3$, compared to applying the gates one by one. The waveform sequence we used is shown schematically in Fig.~\ref{fig:s1}(b).

\begin{table*}[]
	\caption{Experimental parameters of all 12 qubits. $\omega_{01}$ is the idle frequency of qubit and $\eta$ is the anharmonicity. $\omega_{01, opt}$ is the frequency of the qubits at their optimal points. $T_1$, the energy relaxation time, and $T_2^*$, the dephasing time extracted from Ramsey experiment, are measured at idle frequency. $F00$ ($F11$) is the probability of correctly readout of qubit state in $|0\rangle$ ($|1\rangle$) after being well prepared in $|0\rangle$ ($|1\rangle$). $Y/2$ fidelity is characterized by RB at the qubit's idle frequency. }
	\label{table:1}
	\resizebox{\textwidth}{!}{%
		\begin{tabular}{lcccccccccccc}
			\hline
			\textbf{}                 & \textbf{Q1} & \textbf{Q2} & \textbf{Q3} & \textbf{Q4} & \textbf{Q5} & \textbf{Q6} & \textbf{Q7} & \textbf{Q8} & \textbf{Q9} & \textbf{Q10} & \textbf{Q11} & \textbf{Q12} \\ \hline
			$\omega_{01}$ (GHz)        & 4.978      & 4.183      & 5.192      & 4.352      & 5.110      & 4.226      & 5.030      & 4.300      & 5.142      & 4.140       & 4.996       & 4.260       \\
			$\omega_{01, opt}$ (GHz)   & 5.086      & 4.628      & 5.226      & 4.753      & 5.197      & 4.765      & 5.147      & 4.755      & 5.274      & 4.551       & 5.107       & 4.711       \\
			$\eta$ (MHz)               & -248      & -204      & -246      & -203      & -247      & -202      & -246      & -203      & -244      & -203       & -246       & -201       \\
			$T_1$ ($\mu s$)            & 40.1        & 34.7        & 30.8        & 43.2        & 31.8        & 34.3        & 46.5        & 38.1        & 32.2        & 54.6         & 29.6         & 30.3         \\
			$T_2^*$ ($\mu s$)          & 7.9         & 1.5         & 6.3         & 2.4         & 4.9         & 2.7         & 6.8         & 2.3         & 5.1         & 3.5          & 5.9          & 3.0          \\
			$f_{00}$ (\%)                 & 82.8        & 94.4        & 97.9        & 95.8        & 96.1        & 95.8        & 97.2        & 95.4        & 98.5        & 97.1         & 97.7         & 96.5         \\
			$f_{11}$ (\%)                 & 80.0        & 83.8        & 86.7        & 79.5        & 90.9        & 89.7        & 90.8        & 89.6        & 90.1        & 89.2         & 91.1         & 81.7         \\
			$Y/2$ fidelity (\%) & 99.97(11)       & 99.81(6)       & 99.75(22)       & 99.66(11)       & 99.84(6)       & 99.81(8)       & 99.95(7)       & 99.80(6)       & 99.83(7)       & 99.85(22)        & 99.90(9)        & 99.85(12)    \\
			
			\multicolumn{1}{l}{ CZ fidelity (\%)}  & 93.34(35)      &87.41(37)    &92.35(40)       &95.74(39)    &93.78(34)    &93.82(33)    &94.45(32)    &95.64(32)    &95.00(32)       &95.31(31)    &96.42(32)    \\ \hline
		\end{tabular}%
	}
\end{table*}

In order to calibrate the quantum circuit, we first choose the idle point frequencies of the qubits. These frequencies should be as close as possible to the symmetric optimal points listed in Table \ref{table:1} while avoiding large drops in the coherence time caused by TLSs (two-level systems). To reduce $ZZ$ coupling, qubit frequencies alternate between a range of high and low frequencies in a zigzag pattern. Beyond that, qubits in the high and low frequency groups are still mutually detuned by tens of megahertz, to mitigate $XY$ crosstalk.

 After the idle points are chosen, we design the preliminary waveform for the gates. The first step is to find the frequency at which the gate will take place. This frequency is called the operating point. A suitable operating point, one that avoids noticeably high energy-relaxation rates $\Gamma$ caused by TLSs, is initially chosen heuristically. Then, at this operation point, we perform a preliminary control-phase gate and measure the amount of phase shift present in the $|11\rangle$ state. A true CZ gate has a phase shift of $\pi$, but simply measuring the phase shift of a single gate in isolation does not make sense because in our experiment, ten other entangling gates are applied together, and the gates all affect each other. Therefore, when we characterize a single entangling gate, we run the entire three-layer sequence (see Fig. 2(a) in the main text). To isolate the result of a single entangling gate, all other qubits are initialized to the $|0\rangle$ state. This avoids the 12-qubit entanglement that would result from the operation and allows us the characterize the effect of a single entangling gate using quantum process tomography (QPT). QPT allows us to know the phase shift of the gate. By changing the amplitude of the pulse, the strength of the avoided crossing interaction can be controlled, which in turn controls the amount of extra phase accumulated by the $|11\rangle$ state. By choosing the appropriate amplitude, we can make the accumulated phase approximately $\pi$.

 Once the basic shape of the waveform has been designed manually, we implement an iterative process to optimize the waveforms of all gates. The goal is to maximize the gate fidelity, which is defined as the inner product of the desired state and the actual state. The individual entangling gates are optimized by characterizing the gate and adjusting an 8-parameter waveform \cite{Martinis2014} according the Nelder-Mead algorithm. {To minimize the time cost in determining the CZ gate fidelity, we focus on the most significant error sources, which are state leakage from $|11\rangle$ to $|02\rangle$ and phase error in $|++\rangle$. Therefore, in the optimization, the objective function is defined as the combination of state fidelities initialized with $|11\rangle$ and $|++\rangle$. Meanwhile, as the $|++\rangle$ state plays a much more important role in our realization of LC state, the weight of $|++\rangle$ state in objective function is twice as much as $|11\rangle$ state.} Once the fidelity has reached a local maximum, a different gate is optimized, holding all other gate waveforms constant. Optimizing one gate affects the fidelities of the other gates, so we jump from one gate to another, iterating each gate individually, until the fidelities of all the gates become stable.

\begin{figure*}
	\centering
	\includegraphics[width=\textwidth]{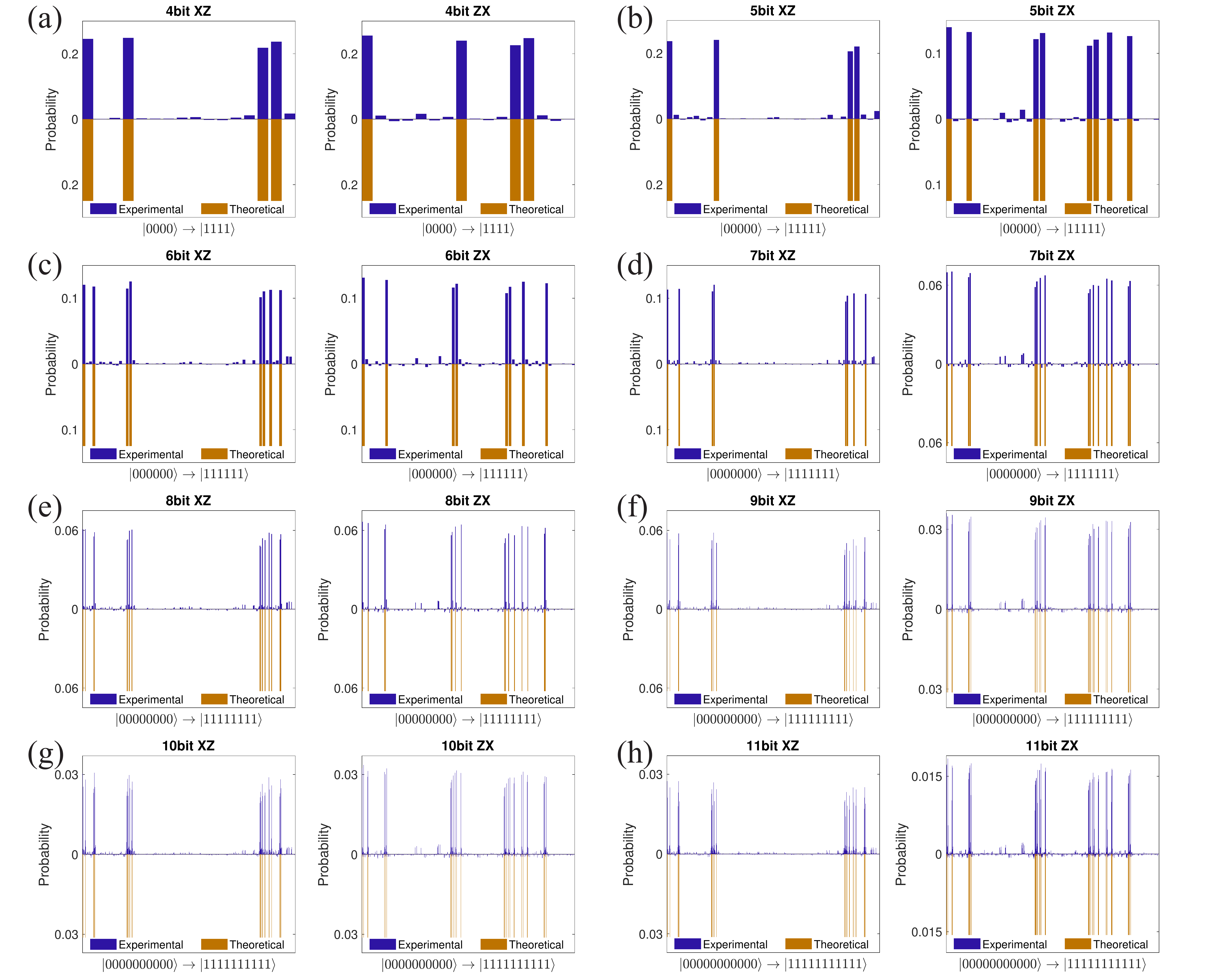}
	\caption{Fig (a)-(h) Experimental and theoretical distribution from 4-qubit to 11-qubit linear cluster state in 12-qubit experiment. The results for two components, $XZ...XZ$ and $ZX...ZX$, are presented separately. }
	\label{fig:s3}
\end{figure*}

After the calibration and optimization, we create LC states with $N$ ranging from 4 to 12. Results from the measured $\sigma_{XZ...XZ}$ and $\sigma_{ZX...ZX}$ components of the LC states from $N=4$ to $11$ are shown in Fig.~\ref{fig:s3}. $N=12$ is shown in the main text.

 \section{Related Experiments}

TLSs in the system cause a big drop in the fidelity of the gates, making them the main obstacle to larger implementations of multi-qubit entanglement. As can be seen in Fig. 2(c) in the main text, the $Q_2$-$Q_3$ CZ gate fidelity is only 0.874 --- clearly lower than the average gate fidelity 0.939. This is what causes the significant decline of the state fidelity that can be observed once the cluster state reaches a size of 11 qubits (seen in Fig. 2(c) in the main text). To our knowledge, this issue is caused by TLSs. There is one coupled to $Q_2$ at $4.34$ GHz with a strength over $20$ MHz and another coupled to $Q_3$ at $4.435$ GHz (see Fig. 1(b) in main text). To realize the CZ gate between $Q_2$ and $Q_3$, $Q_3$ is tuned down to $4.47$ GHz, and $Q_2$ is tuned up to $4.208$ GHz. During this process both qubits come close to the TLSs. 

{To avoid the influence of the TLSs, in our most recent experiment, we warmed up the system to room temperature and cooled it down again. After that, the major TLSs coupled to $Q_2$ and $Q_3$ moved and had less effect in the regime between the qubits' idle points and CZ operating points. As shown in Fig. \ref{fig:s1} (c), in the new experiment, the 12-bit LC state fidelity is about 0.7, significantly higher than the corresponding value 0.55 reported in the main text. Such result supports our understanding that TLS is the most immediate obstacle towards scaling to larger systems.}

\begin{figure*}
	\centering
	\includegraphics[width=0.8\textwidth]{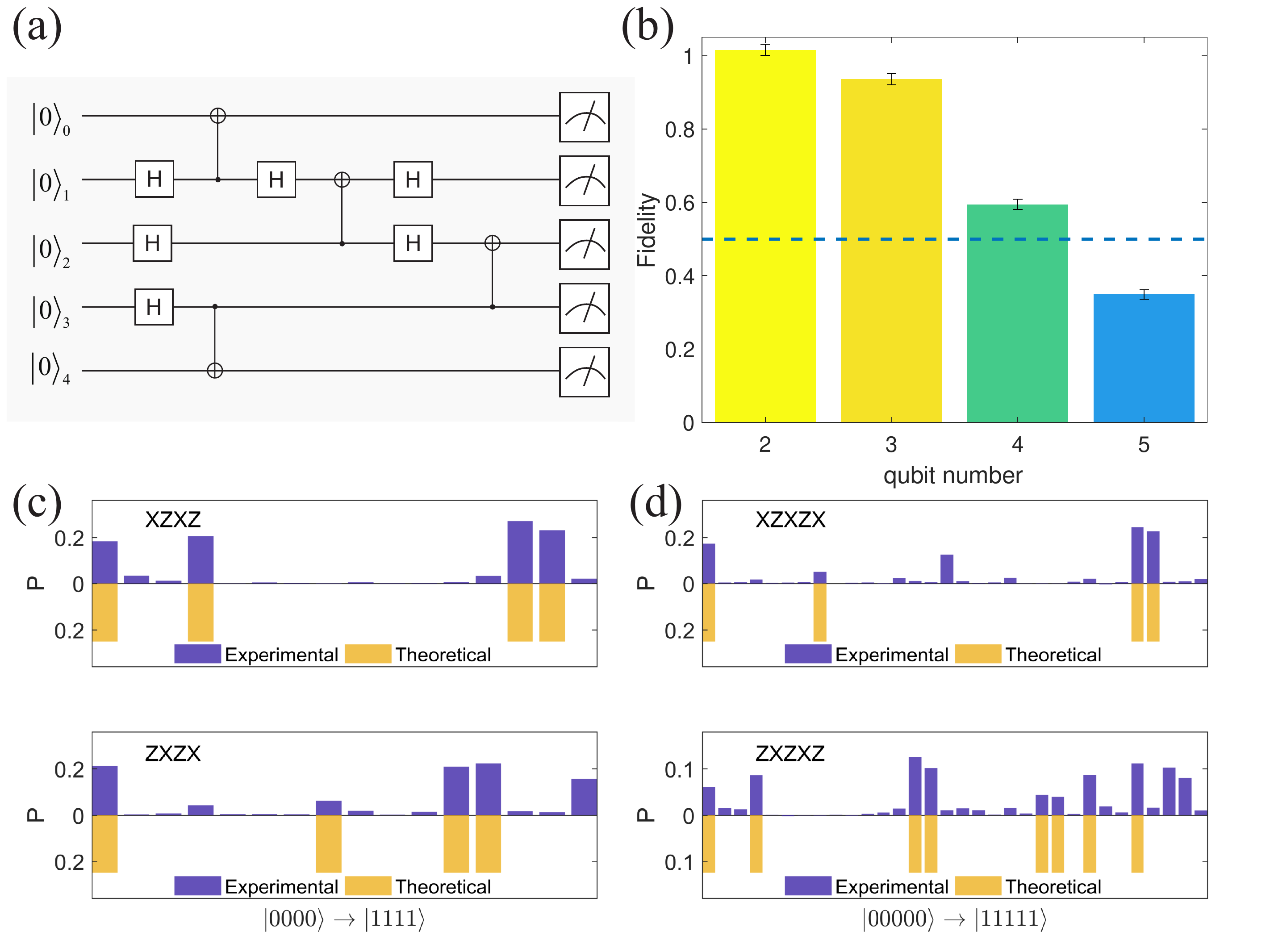}
	\caption{(a) Circuit diagram for 5-qubit linear cluster state generation on IBM-Q system. Different from the circuit used in our 12-qubit experiment, the two-qubit entangling gate supported on the system is CX gate. (b) Linear cluster state fidelities from 2 to 5 qubits. The genuine multi-particle entanglement threshold are plotted in blue dashed line, indicating the existence of genuine multi-particle entanglement for $N \le 4$ qubits. (c) and (d) State distributions of two components for the 4-qubit and 5-qubit LC state, respectively. Experimental and theoretical results are presented in blue and yellow, respectively.}
	\label{fig:s4}
\end{figure*}

In the very recently reported 16 superconducting qubits work on IBM-Q system\cite{16ibm}, the authors stated \blockquote{For a state $\rho$ of a many-body system, for any fixed bipartition AB of the system, if $\rho$ is entangled with respect to the partition AB, then the entanglement of the many-body state $\rho$ can also be examined via its subsystems. That is, if the subsystems are all entangled,  the whole system must be also entangled.} From here, they concluded that there is multipartite entanglement in the prepared linear cluster states on the IBM-Q system. According to this claim, if all the two-qubit reduced density matrices of neighboring qubits under local operations are inseparable, the entire quantum state is also inseparable and hence a multipartite entangled state. Here, we show that this claim is not true due to the limited purity of a quantum state.

Consider the concrete example of a three-qubit mixed state:
\begin{equation}
\begin{aligned}
{\rho _{123}} = {} & \frac{1}{2} | {EPR_{12}}  \rangle \langle {EPR_{12}} | {{0_3}} \rangle  \langle {{0_3}} | \\
				 +  & \frac{1}{2}| {{0_1}} \rangle \langle {{0_1}} | {EPR_{23}} \rangle \langle {EPR_{23}} |
\end{aligned}
\end{equation}
where the $\left| {EPR} \right\rangle  = \left( {\left| {00} \right\rangle  + \left| {11} \right\rangle } \right)/\sqrt 2 $. The reduced density matrix of qubit 1-2 system is produced by tracing out qubit 3. $\left| {0} \right\rangle $ is
$${\rho _{12}} = \frac{2}{3}\left| {EPR_{12}} \right\rangle \left\langle {EPR_{12}} \right| + \frac{1}{3}\left| {{{00}_{12}}} \right\rangle \left\langle {{{00}_{12}}} \right|$$
and similaryly for the qubit 2-3 system $\left| {\text{0}} \right\rangle $ is
$${\rho _{23}} = \frac{2}{3}\left| {EPR_{23}} \right\rangle \left\langle {EPR_{23}} \right| + \frac{1}{3}\left| {{{00}_{23}}} \right\rangle \left\langle {{{00}_{23}}} \right|$$
The entanglement negativity of ${\rho _{{12}}}$ and ${\rho _{{23}}}$ is $\frac{{1}}{{3}}$ , so there is entanglement between qubits 1 and 2 and 2 and 3. According to the criterion in {Ref. \cite{16ibm}}, the whole system must therefore be fully entangled.  However, the system is just a classical mixture of biseparable states $\left| {EPR_{12}} \right\rangle \left\langle {EPR_{12}} | {{0_3}} \right\rangle \left\langle {{0_3}} \right|$  and $\left| {{0_1}} \right\rangle \left\langle {{0_1}} | {EPR_{23}} \right\rangle \left\langle {EPR_{23}} \right|$, which is not genuine three-body entanglement. So, we conclude that a quantum state inseparable with respect to fixed partitions is not necessarily a GME state.

    In August 2018, we tested the LC state generation on IBM 5-qubit system ($ibmqx4$). On that system, the average fidelity of single qubit gates was 0.9985 and the average readout fidelity was 0.9104, according to the website. The circuit for the 5-qubit linear cluster state generation is illustrated in Fig. \ref{fig:s4}(a). The average C-NOT gate fidelity used in the circuit is 0.9519. Fidelities {of LC states} were obtained in the same way as the main text. For each measurement, 8192 readouts were performed. For qubits 2 to 5, the state fidelity was calculated to be 1.0156(157), 0.9355(154), 0.5942(140) and 0.3488(128), respectively (shown in Fig. \ref{fig:s4}(b)). Such results show that we were unable to demonstrate genuine multipartite entanglement for states above 4 qubits. The state distributions for 4-qubit and 5-qubit LC states are shown in Fig. \ref{fig:s4}(c)(d), respectively. It should be noted that in our test, the readout error has been traced out and the readout results of the states have been calibrated based on the directly measured transition matrices in the form of Eq.\ref{transitionmatrix}, thus the readout error is not the main error.

    Reaching larger entanglements is still a crucial task. We believe that the fidelity of the IBM experiment would be even higher if the pulses were calibrated and the quantum circuit optimized. Besides that,  coherence time and tunability of the processor must be sufficiently long and high. Achieving sizable entanglements still needs an overall improvement of the system quality.

	\section{Fidelity Measurement}

The projector for the LC state can be rewritten as a product of stabilizing operators.
\begin{align}
| {L{C_N}} \rangle \langle {L{C_N}} | = \mathop \prod \limits_{i = 1}^N \frac{ {1 + {s_i}}}{2}
\end{align}
But due to the high dimensionality of this operator for $N=12$, it is not practical to measure all 4096 dimensions of this Hilbert space. However, we can define two similar operators $ODD$ and $EVEN$.
\begin{align}
ODD = \mathop \prod \limits_{i \in odd}^N \frac{{1 + {s_i}}}{2}  \\
EVEN = \mathop \prod \limits_{i \in even}^N \frac{{1 + {s_i}}}{2}
\end{align}
These operators are useful for two reasons. First, the expectation values of each of the 64 terms in the expansion of the $ODD$ ($EVEN$) operator can be calculated from the measured distribution of just one operator:  ${\sigma _X}{\sigma _Z}\dots{\sigma _X}{\sigma _Z} $ (${\sigma _Z}{\sigma _X}\dots{\sigma _Z}{\sigma _X}$). Second, the $ODD$ and $EVEN$ operators are related to the LC state projector \cite{witness05,witness09,witness11prl}  by the inequality
\begin{align}
\langle LC \rangle \ge \langle ODD \rangle + \langle EVEN \rangle - 1
\end{align}

\begin{figure}
	\centering
	\includegraphics[width=0.5\textwidth]{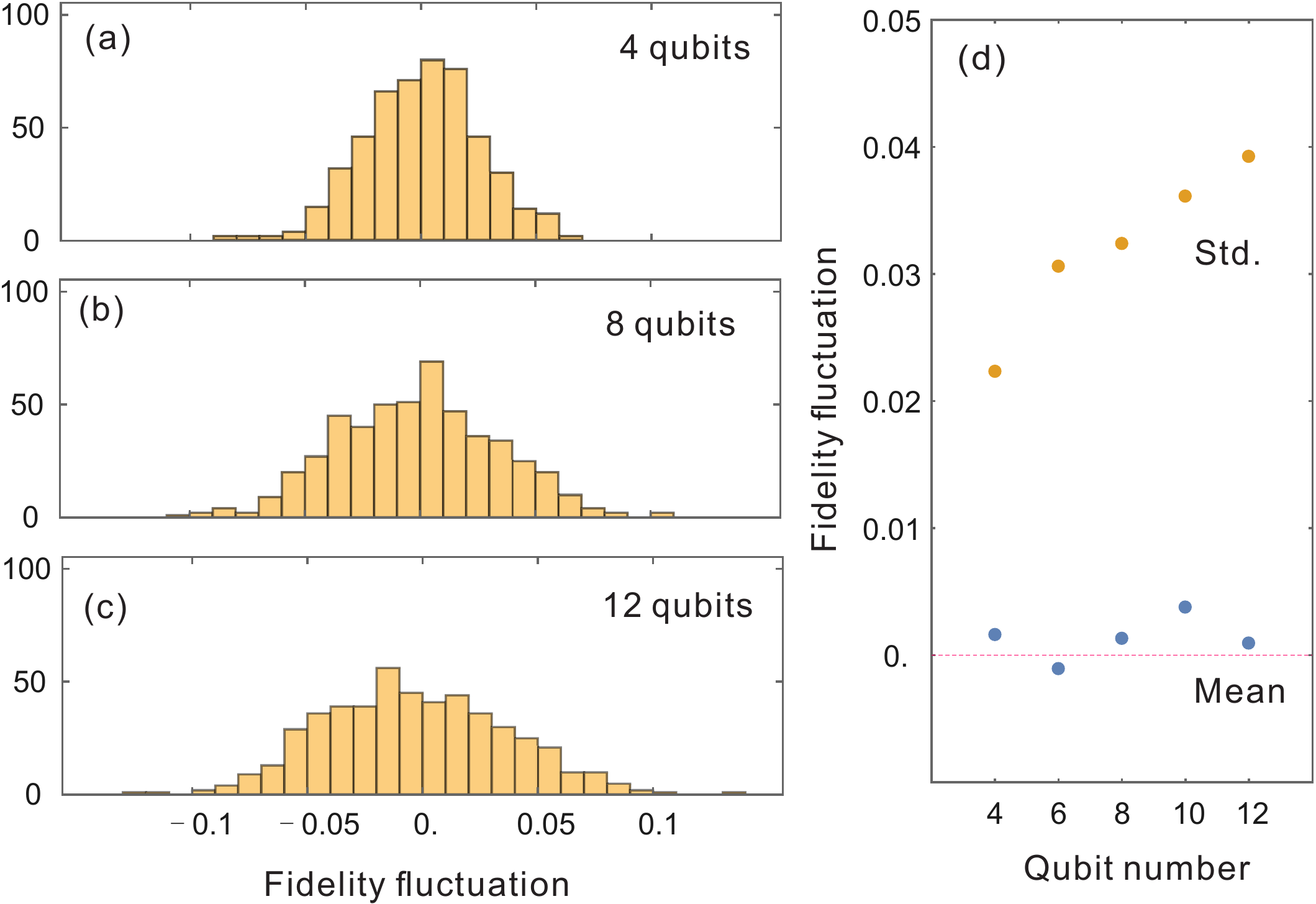}
	\caption{Simulation of distortion of fidelity due to Gaussian fluctuation of transition matrices. We set $f_{00}$ and $f_{11}$ to 0.96 and 0.87, respectively and set the $\delta f_{00}$ and $\delta f_{11}$ to 0.01. (a)(b)(c) The histogram of fidelity distortion for 4, 8 and 12 qubits, respectively. (d) The mean and standard deviation of  fidelity distortion. The mean values are approximately zero. }
	\label{fig:s2}
\end{figure}

To account for our imperfect readout fidelity, the raw distributions, $P_{ZX}^0$ and $P_{XZ}^0$, are multiplied by transition matrices.
\begin{equation}
T_n=\left( {\begin{array}{*{20}{c}}
		{{f_{00}^n}}      &{1 - {f_{11}^n}}    \\
		{1 - {f_{00}^n}}    &    {  {f_{11}^n}    }
		\end{array}} \right)
	\label{transitionmatrix}
\end{equation}
where the diagonal terms represent the probability that the nth qubit, prepared in the $|0\rangle$ ($|1\rangle$) state, will be accurately measured. These values can be found in Table \ref{table:1}. {For multi qubits, we measure the T matrices of different qubits individually and define the multi-qubit T matrix as $T_{i,j}=T_i\otimes T_{i+1}\dots \otimes T_{j}$. With the 12-qubit probabilities $P_r$ defined as a column vector for all 12 qubits, we can correct $P_r$ from the measured probabilities $P_m$ as $P_r=T_{1,12}^{-1}P_m$ \cite{Steffen2006c,Zheng2017}.} Next we define $\vec{\alpha}_{XZ}$ and $\vec{\alpha}_{ZX}$, as the probability distributions resulting from the measurement of a pure $|LC \rangle$ state. These distributions will be compared with our experimental distributions to calculate the lower bound of our state fidelity as follows. A dot product gives the relative distance from each other.
\begin{align}
F &= {\vec{\alpha} _{XZ}} \cdot ({T^{ - 1}} \cdot P_{XZ}^0)  + {\vec{\alpha}_{ZX}} \cdot ({T^{ - 1}} \cdot P_{ZX}^0) - 1 \\
  &= {\vec{\alpha} _{XZ}} \cdot  P_{XZ} + {\vec{\alpha}_{ZX}} \cdot  P_{ZX} - 1
\end{align}

The ensure statistical fluctuations of the system are not responsible for a false-positive result, we reproduce our results with a numerical simulation. Fluctuations of the measured distribution ($\delta F_N$)  and time-dependent fluctuations of the transition matrices ($\delta F_T$) contribute additively to the overall uncertainty of the fidelity bound.
\begin{align}
{\delta F_N}     &= \sqrt {     \frac {   {{\left| {{\vec{\alpha}_{XZ}} \cdot {T^{ - 1}} \cdot \sqrt {p_{XZ}^0} } \right|}^2} + {{\left| {{\vec{\alpha}_{ZX}} \cdot {T^{ - 1}} \cdot \sqrt {p_{ZX}^0} } \right|}^2}  }      {     N     }            }
\end{align}
\begin{align}
\delta {F_T}(t)  &= {} {\vec{\alpha}_{XZ}} \cdot {T^{ - 1}} \cdot (\delta T(t) \cdot {P_{XZ}}) \\
			&+ {\vec{\alpha}_{ZX}} \cdot {T^{ - 1}} \cdot (\delta T(t) \cdot {P_{ZX}})
\end{align}
\begin{equation}
\delta F  = \delta F_N  + \delta F_T
\end{equation}
where the matrix $\delta {T}(t)$ is
\begin{equation}
	\left( {\begin{array}{*{20}{c}}
		{\delta {f_{00}}}&{ - \delta {f_{11}}}\\
		{ - \delta {f_{00}}}&{\delta {f_{11}}}
		\end{array}} \right)
\end{equation}
When $N$ is sufficiently large, fluctuations of $\delta F_N$ can be brought down to an acceptably low rate. And fluctuations of $\delta F_T$ vary on a time scale orders of magnitude lower than the duration of the experiment, so these fluctuations should average away. Our simulations model $f_{00}$ and $f_{11}$ as two Gaussian distributions having means of 0.96 and 0.87, respectively, and both having a standard deviation of 0.01. Fig. \ref{fig:s2} shows the results of the simulation. The mean fluctuations of the fidelity stay near zero as the number of qubits increases. This is consistent with our analysis.
	
\bibliographystyle{apsrev4-1}
%\bibliography{12cluster}
	